\documentclass[aps,pra,twocolumn,superscriptaddress]{revtex4}

\usepackage{bm}

\usepackage{epsfig}
\usepackage{graphicx}
\usepackage{hyperref}
\usepackage{dcolumn}
\usepackage{color}
\usepackage{amsmath, amsthm, amssymb, amsfonts}

\usepackage{xcolor}
\usepackage{bbold}

\newcommand{\ver}{\mathbf{r}}

\newcommand{\vA}{\mathbf{A}}

\newcommand{\vk}{\mathbf{k}}

\begin{document}


\title{
Extracting photoelectron spectra from the time-dependent wave function: Comparison of the projection onto continuum states
and window-operator methods}


\author{B. Feti\' c}
\email[]{benjamin.fetic@gmail.com}
\affiliation{Faculty of Science, University of Sarajevo, Zmaja od Bosne 35, 71000 Sarajevo,
Bosnia and Herzegovina}
\author{W. Becker}
\affiliation{Max-Born-Institut, Max-Born-Str.~2a, 12489 Berlin,
Germany} \affiliation{National Research Nuclear University MEPhI,
Kashirskoe Shosse 31, 115409, Moscow, Russia}
\author{D. B. Milo\v{s}evi\'{c}}
\affiliation{Faculty of Science, University of Sarajevo, Zmaja od
Bosne 35, 71000 Sarajevo, Bosnia and Herzegovina}
\affiliation{Academy of Sciences and Arts of Bosnia and Herzegovina,
Bistrik 7, 71000 Sarajevo, Bosnia and Herzegovina}
\affiliation{Max-Born-Institut, Max-Born-Str.~2a, 12489 Berlin,
Germany}


\date{\today}

\begin{abstract}
Over the last three decades numerous numerical methods for solving the time-dependent Schr\"{o}dinger equation within the single-active
electron approximation have been developed for studying ionization of atomic targets exposed to an intense laser field. In addition,
various numerical techniques for extracting the photoelectron spectra from the time-dependent wave function have emerged. In this paper 
we compare photoelectron spectra obtained by either projecting the time-dependent wave function at the end 
of the laser pulse onto the continuum state having proper incoming boundary condition or by using the 
window-operator method. Our results for three different atomic targets show that the boundary condition imposed onto the continuum states
plays a crucial role for obtaining correct spectra accurate enough to resolve fine details of the interference structures
of the photoelectron angular distribution.
\end{abstract}


\maketitle

\section{Introduction}
The pioneering  work of K. C. Kulander in the late 1980s \cite{kulander,kulander1} has paved the way for the numerical solution of the time-dependent
Schr\"{o}dinger equation (TDSE) to become a very important and powerful tool for studying the laser-atom interaction and related strong-field phenomena. The constant increase in computer power and processor speed of personal computers in the last thirty years has led to the development of numerous numerical methods for solving the TDSE (see, for example, \cite{tong_gs, muller, nurhuda, bengtsson, peng, gordon, telnov}). 
Nowadays, many software codes are available for studying processes such as multiphoton ionization, above-threshold ionization, 
high-order above-threshold ionization, and high-order harmonic generation \cite{qprop, altdse, cltdse, scid-tdse}. 
All these methods have one thing in common, namely the TDSE is solved within the single-active-electron (SAE) approximation for a model 
atom, while the laser-atom interaction is treated in dipole approximation, either using the length or the velocity gauge form 
of the interaction operator. 

Propagation of an initial bound state under the influence of a strong laser field is only one part of 
the problem. Extraction of the physical observables at the end of the laser pulse poses another challenging task. Modern-day 
photoionization experiments designed for recording photoelectron spectra (PES) can be used to simultaneously measure the photoelectron 
kinetic energy and its angular distribution (see, for example, \cite{holo, pad_exp1, pad_exp2}). As the resolution of these experimental
techniques increased, the theoretical calculation of highly accurate PES from  \emph{ab initio} methods such as numerical solution of the 
TDSE became essential in order to distinguish different mechanisms that play a role in a photoionization process.

Formal exact PES for a one-electron photoionization process can be calculated by projection of the time-dependent wave function at the end of the laser pulse onto the continuum states of the field-free Hamiltonian. We call this method the PCS (Projection onto Continuum States) method.
For long laser pulses at near-infrared wavelengths and moderate intensities the photoelectron can travel very far away from the origin.
In order to include the fastest photoelectrons the volume within which the wave function is simulated has to be very large. 
Another deficiency of the PCS method is that the continuum states, onto which we project the solutions
of the TDSE at the end of the laser pulse, are analytically known only for the pure Coulomb potential, while for non-Coulomb potentials
they have to be obtained numerically. That is why many approximative methods for extracting PES with no need to calculate the continuum 
states have emerged in the last three decades. One of the earliest methods used for extracting the PES from the time-dependent wave 
function is the so-called window-operator (WO) method \cite{wop}. It has been successfully used in the past for PES calculations for
atomic targets exposed to a strong laser field \cite{wop_app}. Recently, the WO method has also been used for studying high-order
above-threshold ionization of the H$_2^+$ molecular ion \cite{fetic_mhati}. 
There is also the so-called tSURFF method \cite{tsurff},  which is designed to replace the projection onto continuum states
with a time integral of the outer-surface flux, allowing one to use much a smaller simulation volume. An extension of the tSURFF method called 
iSURF method \cite{morales} has also been used for calculating PES. Another way of calculating PES without explicit calculation of the
continuum states is to propagate the wave function  under the influence of the field-free Hamiltonian for some period of time after the 
laser pulse has been turned off, so that even the slowest parts of the wave function have reached the asymptotic zone \cite{madsen}. However, for neutral atomic targets
this method requires a large spatial grid to include the part of the wave function associated with 
the fastest photoelectrons. 

From a numerical point of view, the above-mentioned approximative methods may be appealing since they are less time consuming than the
exact PCS method. However, they can mask some fine details in the PES due to neglecting the nature of the continuum
state associated with a photoelectron. Therefore, approximative methods used for extracting PES from the wave function have to be checked 
for consistency by comparing with the exact method. In this paper we compare the results obtained using the exact PCS method with those
obtained with the WO method.

This paper is organized as follows. In Sec.~\ref{sec:num} we first describe our numerical method for solving the Schr\"{o}dinger equation. 
Next, we introduce the method of extracting PES from the time-dependent wave function using the method of projecting onto continuum states 
and the window-operator method. In Sec.~\ref{sec:results} we present our results for PES obtained by these two methods. We compare 
results for three different targets, fluorine negative ions and hydrogen and argon atoms, modeled by different types of the binding 
potential. Finally, we summarize our results and give conclusions in Sec.~\ref{sec:sum}.
Atomic units (a.u.; $\hbar=1$, $4\pi\varepsilon_0=1$, $e=1$, and $m_e=1$) are used throughout the paper, unless otherwise stated.

\section{Numerical methods}\label{sec:num}
\subsection{Method of solving the Schr\"{o}dinger equation}\label{subsec:tdse}
We start by solving the stationary Schr\"{o}dinger equation for an arbitrary spherically symmetric binding potential $V(\ver)=V(r)$
in spherical coordinates:
\begin{equation}
H_0\psi(\ver)=E\psi(\ver),\quad H_0=-\frac{1}{2}\nabla^{2}+V(r).
\end{equation}
We are looking for solutions in the form
\begin{equation}
\psi_{n\ell m}(\ver)=\frac{u_{n\ell}(r)}{r}Y_\ell^m(\Omega),\quad \Omega\equiv (\theta,\varphi),
\end{equation}
where the  $Y_{\ell}^{m}(\Omega)$ are spherical harmonics. The radial function $u_{n\ell}(r)$ is a solution of the radial Schr\"{o}dinger 
equation:
\begin{equation}
H_\ell(r)u_{n\ell}(r) =  E_{n\ell}u_{n\ell}(r), \label{tdse:rad}
\end{equation}
\begin{equation}
H_\ell(r)=-\frac{1}{2}\frac{d^{2}}{dr^{2}}+V(r)+\frac{\ell(\ell+1)}{2r^{2}},\label{tdse:rad1}
\end{equation}
where $n$ is the principal quantum number and $\ell$ is the orbital quantum number. For bound states with the energy $E_{n\ell}<0$ the
corresponding radial wave function $u_{n\ell}(r)$ has to obey the boundary conditions $u_{n\ell}(0)=0$ and $u_{n\ell}(r) \to 0$ for 
$r\to\infty$. The radial equation (\ref{tdse:rad}) is solved numerically in the interval $[0,r_{\max}]$ by expanding the radial 
function into the B-spline basis set as
\begin{equation}
u_{n\ell}(r) = \sum_{j=2}^{N-1}c_{j}^{n\ell}B_{j}^{(k_{s})}(r),\label{tdse:bsp}
\end{equation}
where $N$ represents the number of B-spline functions in the domain $[0,r_{\max}]$ and $k_s$ is the order of the B-spline function. 
All results presented in this paper have been obtained using the order $k_s=10$ and for simplicity we omit it in further expressions. 
Since we require that the radial function vanishes at the boundary, we exclude the first and the last B-spline function
in the expansion (\ref{tdse:bsp}). For more details on the properties of the B-spline basis, see \cite{Bachau}.

Inserting (\ref{tdse:bsp}) into (\ref{tdse:rad}), multiplying the obtained equation with $B_{i}(r)$, and integrating over the radial 
coordinate for fixed orbital quantum number $\ell$, we obtain a generalized eigenvalue problem in the form of a matrix equation:
\begin{equation}
\mathbf{H}_{0}^{\ell}\mathbf{c}^{n\ell}=E\mathbf{S}\mathbf{c}^{n\ell},\label{tdse:eigen}
\end{equation}
where
\begin{equation}
\left(\mathbf{H}_0^\ell\right)_{ij}=\int_0^{r_{\max}}B_i(r)H_\ell(r)B_j(r)dr,
\end{equation}
\begin{equation}
(\mathbf{S})_{ij}=\int_{0}^{r_{\max}}B_{i}(r)B_{j}(r)dr.
\end{equation}
The overlap matrix $\mathbf{S}$ originates from the fact that the B-spline functions do not form an orthogonal basis set. All integrals
involving B-spline functions are calculated with the Gauss-Legendre quadrature rule. Using standard diagonalization procedure for
solving (\ref{tdse:eigen}) we obtain the ground-state energy and the corresponding eigenvector, which is used as an initial state in the TDSE.

In order to describe the laser-atom interaction we numerically solve the time-dependent Schr\"{o}dinger equation:
\begin{equation}
i\frac{\partial\Psi(\ver,t)}{\partial t}=\left[H_0+V_I(t)\right]\Psi(\ver,t),\label{tdse}
\end{equation}
where $V_I(t)$ is the interaction operator in the dipole approximation and velocity gauge. We assume that the laser field is linearly 
polarized along the $z$ axis, so that the interaction operator can be written as 
\begin{eqnarray}
V_I(t)=-i\vA(t)\cdot\mathbf{\nabla}=-iA(t)\left(\cos\theta\frac{\partial}{\partial r}-\frac{\sin\theta}{r}\frac{\partial}{\partial\theta}\right),
\end{eqnarray}
where $A(t)=-\int^{t}E(t')dt'$ and $E(t)$ is the electric field given by
\begin{eqnarray}
E(t)=E_0\sin^2\left(\frac{\omega t}{2N_c}\right)\cos(\omega t),\quad t\in[0,T_p],
\end{eqnarray}
where $\omega=2\pi/T$ is the laser-field frequency and $T_p=N_cT$ is the pulse duration, with $N_c$ the number of
optical cycles. The amplitude $E_0$ is related to the intensity $I$ of the laser field by the relation $E_0=\sqrt{I/I_A}$ where 
$I_A=3.509\times 10^{16}~\text{W}/\text{cm}^{2}$ is the atomic unit of intensity.

The TDSE is solved by expanding the time-dependent wave function in the basis of B-spline functions and spherical harmonics:
\begin{equation}
\Psi(r,\Omega, t) = \sum_{j=2}^{N-1}\sum_{\ell=0}^{L-1} c_{j\ell}(t)\frac{B_{j}(r)}{r}Y_{\ell}^{m_0}(\Omega),
\label{tdse:expan}
\end{equation}
where the expansion coefficients $c_{j\ell}(t)$ are time-dependent. For a linearly polarized laser field, the magnetic quantum number is 
constant and we set it equal to $m_0=0$.
Inserting the expansion (\ref{tdse:expan}) into (\ref{tdse}), multiplying the obtained result by $B_{i}(r)Y_{\ell'}^{m_{0}*}(\Omega)/r$, and integrating 
over the spherical coordinates, we obtain the TDSE in the form of the following matrix equation:
\begin{eqnarray}
i(\mathbf{S}\otimes\mathbb{1}_{\ell})\frac{d\mathbf{c}(t)}{dt}=\left[\mathbf{H}_{0}^{\ell}\otimes\mathbb{1}_{\ell}
-iA(t)\mathbf{W}_I\right]\mathbf{c}(t),\label{tdse:matrix}
\end{eqnarray}
where $\mathbb{1}_{\ell}$ is the identity matrix in $\ell$-space and
\begin{eqnarray}
\mathbf{c}(t) &=& 
\big[(c_{20},\dots, c_{N-10}),(c_{21},\dots, c_{N-11}),\nonumber\\ &~&\dots,(c_{2L-1},\dots, c_{N-1L-1})\big]^{T},
\end{eqnarray}
is a time-dependent vector. The matrices $\mathbf{S}$ and $\mathbf{H}_{0}^{\ell}$ are diagonal in $\ell$-space while the matrix
$\mathbf{W}_{I}$ couples the $\ell-1$ and $\ell+1$ $\ell$-block:
\begin{eqnarray}
( \mathbf{W}_{I})_{ij}^{\ell'\ell} &=&  (\mathbf{Q})_{ij}\left[\ell c_{\ell-1}^{m_0}
\delta_{\ell',\ell-1} -
(\ell+1)c_{\ell}^{m_0}\delta_{\ell',\ell+1}\right]
\nonumber\\&& +(\mathbf{P})_{ij} \left[c_{\ell-1}^{m_0}\delta_{\ell',\ell-1} + c_{\ell}^{m_0}\delta_{\ell',\ell+1}\right],
\end{eqnarray}
where
\begin{eqnarray}
c_\ell^{m_0}&=&\sqrt{\frac{(\ell+1)^{2}-m_0^2}{(2\ell+1)(2\ell+3)}},\\
(\mathbf{Q})_{ij}&=&\int_{0}^{r_{\max}}\frac{B_{i}(r)B_{j}(r)}{r}dr,\\
(\mathbf{P})_{ij}&=&\int_{0}^{r_{\max}}B_{i}(r) \frac{dB_{j}(r)}{dr}dr.
\end{eqnarray}
Since the matrix $\mathbf{W}_I$ couples only the $\ell-1$ and the $\ell+1$ $\ell$-block, it can be decomposed in a sum of mutually commuting matrices 
\begin{eqnarray}
\mathbf{W}_I = \sum_{\ell=0}^{L-2}\left(\mathbf{P}\otimes\mathbf{L}_{\ell m_0} + \mathbf{Q}\otimes\mathbf{T}_{\ell m_0}\right),
\end{eqnarray}
where
\begin{eqnarray}
\mathbf{L}_{\ell m_0}&=&c_{\ell}^{m_0}\left(\begin{array}{cc}
          0 &  1\\
          1 & 0
         \end{array}\right),\\
         \mathbf{T}_{\ell m_0} &=& 
         (\ell+1)c_{\ell}^{m_0}
         \left(\begin{array}{cc}
          0 & 1 \\
          -1 & 0
         \end{array}\right),
\end{eqnarray}
are effectively $2\times2$ matrices acting upon the vector $[\mathbf{c}_{\ell}, \mathbf{c}_{\ell+1}]^{T}=[(c_{2l},\dots, c_{N-1l}),(c_{2\ell+1},\dots, c_{N-1\ell+1})]^{T}$.

The formal solution of the matrix equation (\ref{tdse:matrix}) can be written as
\begin{eqnarray}
\mathbf{c}(t+\Delta t) &=& \exp\bigg\{-i(\mathbf{S}^{-1} \otimes\mathbb{1}_{\ell})\nonumber\\ &~&\times \int_{t}^{t+\Delta t}
\left[\mathbf{H}_{0}\otimes\mathbb{1}_{\ell} - iA(t')\mathbf{W}_{I}\right]dt'\bigg\} \mathbf{c}(t).\nonumber\\
\end{eqnarray}
The evolution of the inital wave function is described by the same numerical recipe as in \cite{qprop}, but without using finite difference
expressions. Our final expression for this time evolution is
\begin{eqnarray}
\displaystyle\mathbf{c}(t+\Delta t) &=& \prod_{l=L-2}^{0}\Bigg[ \frac{\mathbf{S}\otimes\mathbb{1}_{\ell}-
\frac{\Delta t}{4}A(t+\Delta t)\mathbf{P}\otimes \mathbf{L}_{\ell m_0}}
{\mathbf{S}\otimes\mathbb{1}_{\ell}+\frac{\Delta t}{4}A(t+\Delta t)\mathbf{P}\otimes \mathbf{L}_{\ell m_0}}\nonumber\\
&~&\times \frac{\mathbf{S}\otimes\mathbb{1}_{\ell}- \frac{\Delta t}{4}A(t+\Delta t)\mathbf{Q}\otimes\mathbf{T}_{\ell m_0}}
{\mathbf{S}\otimes\mathbb{1}_{\ell}+\frac{\Delta t}{4}A(t+\Delta t) \mathbf{Q}\otimes \mathbf{T}_{\ell m_0}} \Bigg]
\nonumber\\ &~&\times \prod_{\ell=0}^{L-1} \frac{(\mathbf{S}-i\frac{\Delta t}{2}\mathbf{H}_{0}^{\ell})\otimes\mathbb{1}_{\ell}}
{(\mathbf{S}+i\frac{\Delta t}{2}\mathbf{H}_{0}^{\ell})\otimes\mathbb{1}_{\ell}} \nonumber \\
&~& \times \prod_{\ell=0}^{L-2} \Bigg[ \frac{\mathbf{S}\otimes\mathbb{1}_{\ell}-\frac{\Delta t}{4}A(t)\mathbf{Q}\otimes \mathbf{T}_{\ell m_0}}{\mathbf{S}\otimes\mathbb{1}_{\ell}
+ \frac{\Delta t}{4}A(t)\mathbf{Q}\otimes \mathbf{T}_{\ell m_0}}\nonumber\\&~&\times
\frac{\mathbf{S}\otimes\mathbb{1}_{\ell}-\frac{\Delta t}{4}A(t)\mathbf{P}\otimes \mathbf{L}_{\ell m_0}}
{\mathbf{S}\otimes\mathbb{1}_{\ell}+\frac{\Delta t}{4}A(t)\mathbf{P}\otimes \mathbf{L}_{\ell m_0}}
 \Bigg]\mathbf{c}(t).
\end{eqnarray}

\subsection{Extracting the photoelectron spectra from the time-dependent wave function}
The photoelectron spectra can be extracted from the time-dependent wave function $\Psi(\ver, t)$ at the end of the laser pulse by projecting it 
onto the continuum states having the momentum $\vk=(k,\Omega_\vk)$, $\Omega_\vk\equiv(\theta_\vk,\varphi_\vk)$. These continuum states are 
solutions of the stationary Schr\"{o}dinger equation for an electron moving in a spherically symmetric potential $V(r)$. There are two
linearly independent continuum states labeled $\Phi_{\vk}^{(+)}(\ver)$ and $ \Phi_{\vk}^{(-)}(\ver)$, which satisfy 
different boundary conditions at large distance from the atomic target:
\begin{equation}
\Phi_{\vk}^{(\pm)}(\ver)\xrightarrow{r\to\infty} (2\pi)^{-3/2}\left(e^{i\vk \cdot \ver}
+f^{(\pm)}(\theta_\vk)\frac{e^{\pm ikr}}{r}\right),
\end{equation}
where $f^{(\pm)}(\theta_\vk)$ is the usual scattering amplitude. The solutions $\Phi_{\vk}^{(+)}(\ver)$ represent continuum states that
obey the so-called outgoing boundary condition whereas the solutions 
$\Phi_{\vk}^{(-)}(\ver)$ represent continuum states that obey the so-called incoming 
boundary condition. The difference between these two continuum states becomes manifest in the time dependence of their corresponding wave 
packets as shown in \cite{roman}. Here we only give the main result. Namely, a long time after the interaction with the target, the
continuum states $\Phi_{\vk}^{(+)}(\ver)$ and $\Phi_{\vk}^{(-)}(\ver)$ behave as follows:
\begin{eqnarray}
\Phi_{\vk}^{(+)}(\ver,t)&\xrightarrow{t\to\infty}& (2\pi)^{-3/2} e^{i(\vk \cdot \ver-E_\vk t)} 
+ \text{a scattering wave},\nonumber \\ \Phi_{\vk}^{(-)}(\ver,t)&\xrightarrow{t\to\infty}& (2\pi)^{-3/2} e^{i(\vk \cdot \ver-E_\vk t)}.
\end{eqnarray}
In an ionization experiment, the electron liberated by ionization winds up in a quantum state having linear momentum 
$\vk$. Therefore, the continuum state $\Phi_{\vk}^{(-)}(\ver)$ is suitable for describing an ionization experiment while the continuum 
state $\Phi_{\vk}^{(+)}(\ver)$ is employed for a collision experiment. For more detailed analysis and discussion, see \cite{starace}.

Both continuum states can be written as partial wave expansions:
\begin{equation}
\Phi_{\vk}^{(\pm)}(\ver) = \sqrt{\frac{2}{\pi}}\frac{1}{k}\sum_{\ell,m} i^{\ell}e^{\pm i\Delta_{\ell}}\frac{u_{\ell}(k,r)}{r}
Y_{\ell}^{m}(\Omega)Y_{\ell}^{m*}(\Omega_\vk),\label{cont_st}
\end{equation}
where $\Delta_{\ell}$ is the scattering phase shift of the $\ell$th partial wave.  The radial function $u_{\ell}(k,r)$ is a solution of the radial 
Schr\"{o}dinger equation (\ref{tdse:rad}) for fixed orbital quantum number and kinetic energy $E_\vk=k^{2}/2$. The continuum states  
(\ref{cont_st}) are normalized on the momentum scale, i.e., $\langle  \Phi_{\vk'}^{(\pm)}| \Phi_{\vk}^{(\pm)}\rangle = \delta(\vk'-\vk)$.

For the pure Coulomb potential $V(r) = -Z/r$, the scattering phase shift $\Delta_\ell$ is equal to the Coulomb phase shift 
$\sigma_\ell=\arg\Gamma(\ell+1 + i \eta)$, with $\eta = -Z/k$ the Sommerfeld parameter. The radial function $u_\ell(k,r)$ is given by 
the regular Coulomb function $u_\ell(k,r)=F_\ell(\eta,kr)$, which is known in analytical form.
Coulomb functions $F_{\ell}(\eta,kr)$ and corresponding
phase shifts $\sigma_{\ell}$ are calculated using a subroutine from \cite{peng1}.

For the modified Coulomb potential 
\begin{equation}
V(r) = -\frac{Z}{r} + V_{s}(r),
\end{equation}
the scattering phase shift $\Delta_\ell$ is the sum of the Coulomb phase shift $\sigma_\ell$ and the phase shift $\hat{\delta}_\ell$ due to the
presence of the short-range potential $V_s(r)$. In this case, the radial equation is solved numerically by the Numerov method in the interval 
$r\in [0, r_0]$, where $r_0$ is the chosen size of the spherical box, and the phase shift $\hat{\delta}_\ell$ is obtained by matching the 
numerical solution $u_\ell(k,r)$ to the known asymptotic solution \cite{joachain}:
\begin{equation}
\mathcal{N}u_{\ell}(k, r) = \cos\hat{\delta}_{\ell}F_{\ell}(\eta,kr) + \sin\hat{\delta}_{\ell}G_{\ell}(\eta,kr), \label{matching}
\end{equation}
where $G_{\ell}(\eta,kr)$ is the irregular Coulomb function and $\mathcal{N}$ is a normalization constant. To avoid having to calculate 
derivatives, the phase shift $\hat{\delta}_\ell$ is obtained by matching at two different points $r_1$ and $r_2$ close to the boundary $r_0$:
\begin{equation}
\tan\hat{\delta}_{\ell} = \frac{\kappa F_{\ell}(\eta,kr_{2}) - F_{\ell}(\eta, kr_{1})} {G_{\ell}(\eta,kr_{1}) -\kappa G_{\ell}(\eta, kr_{2})},\quad 
\kappa = \frac{u_{\ell}(k,r_{1})}{u_{\ell}(k, r_{2})}.
\end{equation}

For a pure short-range potential $V(r) = V_{s}(r)$~($\eta=0$), the Coulomb functions $F_{\ell}(\eta,kr)$ and 
$G_{\ell}(\eta,kr)$ must be replaced by the spherical Bessel function $j_{\ell}(kr)$ and the spherical Neumann function $n_{\ell}(kr)$:
\begin{equation}
F_{\ell}(0,kr) = krj_{\ell}(kr), \quad G_{\ell}(0,kr) = -krn_{\ell}(kr).
\end{equation}
The spherical Bessel and Neumann functions and the Coulomb functions are calculated using a subroutine from \cite{coul90}.
After obtaining the phase shift $\hat{\delta}_{\ell}$, the numerical solution $u_{\ell}(k,r)$ is normalized according to (\ref{matching}).

The probability of finding the electron at the end of the laser pulse in a continuum state with the 
momentum $\vk = (k,\Omega_{\vk})$ is given by 
\begin{equation}
P(k, \Omega_{\vk}) = \frac{d^{3}P}{k^{2} dk d\Omega_{\vk}} = \left|\langle \Phi_{\vk}^{(-)} | \Psi(T_{p})\rangle\right|^{2}.\label{pad_1}
\end{equation}
Inserting (\ref{cont_st}) and (\ref{tdse:expan}) into (\ref{pad_1}) we obtain the expression 
\begin{equation}
P(k, \Omega_{\vk})= \frac{2}{\pi}\frac{1}{k^{2}}\Big|\sum_{i,\ell}c_{i\ell}(T_{p})(-i)^{\ell}e^{i\Delta_{\ell}}  
Y_{\ell}^{m_0}(\Omega_{\vk})I_{i\ell}(k)\Big|^{2}, \label{prob}
\end{equation}
where we have introduced the integral
\begin{eqnarray}
I_{i\ell}(k)&=& \int_{0}^{r_{0}}u_{\ell}(k,r)B_{i}(r)dr 
+\int_{r_{0}}^{r_{\max}}\Big[\cos\hat{\delta}_{\ell}F_{\ell}(\eta,kr)
\nonumber\\ &~& + \sin\hat{\delta}_{\ell}G_{\ell}(\eta,kr)\Big] B_{i}(r)dr.
\end{eqnarray}
The photoelectron angular distribution (PAD), i.e., the probability $P(E_\vk,\theta_\vk)$ of detecting the electron with kinetic
energy $E_\vk$ emitted in the direction $\theta_\vk$, is given by replacing $k = \sqrt{2E_\vk}$ in (\ref{pad_1}) and integrating over $\varphi_\vk$:
\begin{eqnarray}
P(E_\vk,\theta_\vk)&=&\frac{d^2P}{\sin\theta_\vk dE_\vk d\theta_\vk}\nonumber\\ &=&\frac{1}{\pi\sqrt{2E_\vk}}\Big|\sum_{i,\ell}c_{i\ell}(T_p)(-i)^{\ell}
e^{i\Delta_{\ell}}\nonumber\\&~&\times\sqrt{2l+1}P_{\ell}^{m_0}(\cos\theta_\vk)I_{i\ell}(k)\Big|^{2},\label{pad_2}
\end{eqnarray}
where $P_{\ell}^{m_0}(\cos\theta_\vk)$ are associated Legendre polynomials.

\subsection{Window-operator method}
Obtaining the photoelectron angular distribution by projecting onto continuum states can be a challenging task since the continuum states are 
highly oscillatory functions. Therefore, the numerical integration has to be done with high precision and stability to get the
photoelectron spectra with an accuracy of a few orders of magnitude. This is especially true for non-Coulomb potentials since in this case the continuum states
must be obtained numerically. In this section we present the implementation of the WO method, which can be used for the extraction of the PES 
without the need to calculate the continuum states. 

The WO method is based on the projection operator $W_{\gamma}(E_\vk)$ defined by
\begin{equation}
 W_{\gamma}(E_\vk) = \frac{\gamma^{2^{n}}}{(H_{0}-E_\vk)^{2^{n}} + \gamma^{2^{n}}},\label{wo}
\end{equation}
which extracts the component $|\chi_{\gamma}(E_\vk)\rangle$ of the final wave vector $|\Psi(T_{p})\rangle$ that contributes to energies 
within the bin of the width $2\gamma$, centered at $E_\vk$:
\begin{equation}
|\chi_{\gamma}(E_\vk)\rangle = W_{\gamma}(E_\vk)|\Psi(T_{p})\rangle.\label{wo_eq}
\end{equation}
We set $n=3$ and expand the wave vector into the basis (\ref{tdse:expan}):
\begin{equation}
\chi_{\gamma}(E_\vk, r,\Omega) = \sum_{i=2}^{N-1}\sum_ {\ell=0}^{L-1}b_{i\ell}^{(\gamma)}(E_\vk)\frac{B_{i}(r)}{r}Y_{\ell}^{m_0}(\Omega).
\end{equation}
To obtain the coefficients $b_{i\ell}^{(\gamma)}(E_\vk)$ we solve Eqn.~(\ref{wo_eq}) by factorizing (\ref{wo}) \cite{qprop} and transforming it into a series of matrix equations:
\begin{eqnarray}
 &~& \mathbb{1}_{\ell}\otimes\left[\mathbf{H}_{0}^{\ell}-\mathbf{S}(E_\vk-\gamma e^{i\nu_{34}})\right]
 \left[\mathbf{H}_0^\ell-\mathbf{S}(E_\vk+\gamma e^{i\nu_{34}})\right]\mathbf{b}_{1}^{(\gamma)} \nonumber\\
 &~& =  \gamma^{2^{3}}\mathbb{1}_{\ell}\otimes\mathbf{S}\mathbf{c}(T_{p}),\nonumber\\ 
  &~& \mathbb{1}_{\ell}\otimes\left[\mathbf{H}_{0}^{\ell}-\mathbf{S}(E_\vk-\gamma e^{i\nu_{33}})\right]
 \left[\mathbf{H}_{0}^{\ell}-\mathbf{S}(E_\vk+\gamma e^{i\nu_{33}})\right]\mathbf{b}_{2}^{(\gamma)} \nonumber\\
 &~& =  \mathbb{1}_{\ell}\otimes\mathbf{S}\mathbf{b}_{1}^{(\gamma)},\nonumber\\
   &~& \mathbb{1}_{\ell}\otimes\left[\mathbf{H}_{0}^{\ell}-\mathbf{S}(E_\vk-\gamma e^{i\nu_{32}})\right]
 \left[\mathbf{H}_{0}^{\ell}-\mathbf{S}(E_\vk+\gamma e^{i\nu_{32}})\right]\mathbf{b}_{3}^{(\gamma)} \nonumber\\
 &~& =  \mathbb{1}_{\ell}\otimes\mathbf{S}\mathbf{b}_{2}^{(\gamma)},\nonumber\\
   &~& \mathbb{1}_{\ell}\otimes\left[\mathbf{H}_{0}^{\ell}-\mathbf{S}(E_\vk-\gamma e^{i\nu_{31}})\right]
 \left[\mathbf{H}_{0}^{\ell}-\mathbf{S}(E_\vk+\gamma e^{i\nu_{31}})\right]\mathbf{b}^{(\gamma)} \nonumber\\
 &~& =  \mathbb{1}_{\ell}\otimes\mathbf{S}\mathbf{b}_{3}^{(\gamma)},
\end{eqnarray}
where $\nu_{3j}= (2j-1)\pi/2^{3}$. After obtaining $\mathbf{b}^{(\gamma)}$, the probability of finding the electron with the energy $E_\vk$ is calculated as
\begin{eqnarray}
 P_{\gamma}(E_\vk) &=& \int dV \chi_{\gamma}^{*}(E_\vk,r, \Omega)\chi_{\gamma}(E_\vk,r, \Omega)\nonumber\\
 &=&\int d\Omega dr P_{\gamma}(E_\vk, r,\Omega),
\end{eqnarray}
where
\begin{eqnarray}
 P_{\gamma}(E_\vk, r,\Omega) = 
 \Bigg| \sum_{i=2}^{N-1}\sum_{\ell=0}^{L-1}
 b_{i\ell}^{(\gamma)}(E_\vk)B_{i}(r)Y_{\ell}^{m_0}(\Omega) \Bigg|^{2}.
\end{eqnarray}
Now we make the assumption that the solid-angle element $d\Omega$ in position space  is approximately
equal to the solid-angle element $d\Omega_\vk$ in momentum space  (for details, see \cite{deGruyter}). This means that information about the probability distribution in energy and in angle
is obtained by integrating $P_{\gamma}(E_\vk, r,\Omega_{\vk})\approx P_{\gamma}(E_\vk, r,\Omega)$ over the radial coordinate.
In
this case we define the probability $P_{\gamma}(E_\vk,
\Omega_{\vk}) = P_{\gamma}(E_\vk, \theta_\vk )/(2\pi)$ which is equal, up
to a constant factor, to the PAD, Eq.~(\ref{pad_2}).

\section{Results and Discussion}
\label{sec:results}
\begin{figure}[b!]
\vspace{1.3cm}
\centering
\includegraphics[scale=0.45]{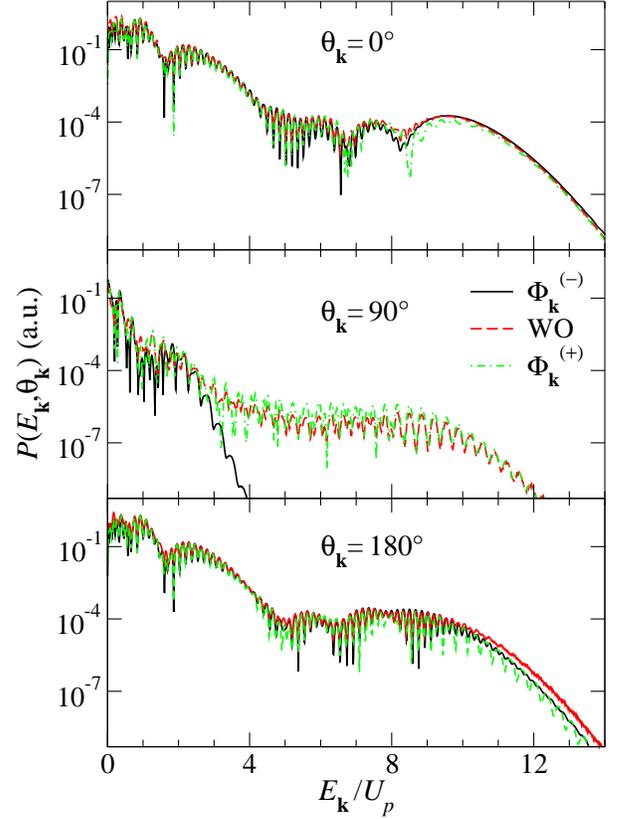}
\caption{The differential detachment probabilities of F$^-$ ions for emission of electrons 
in the directions $\theta_\vk=0^{\circ}$, $90^{\circ}$, and $180^{\circ}$, as functions
of the photoelectron energy in units of the ponderomotive energy $U_p$, for the following
laser-field parameters: $I=1.3\times 10^{13}~\text{W}/\text{cm}^{2}$, 
$\lambda =1800~\text{nm}$, and $N_c=6$. The results are obtained
by projecting the time-dependent wave function $\Psi(T_{p})$ onto the 
$\Phi_{\vk}^{(-)}$ states (black solid line) and $\Phi_{\vk}^{(+)}$ states
(green dot-dashed line) and using the WO method with $\gamma = 2\times 10^{-3}$
(red dashed line).} \label{results:f-_pad_vs_wop}
\end{figure}

\begin{figure}[t!]
\centering
\includegraphics[scale=0.45]{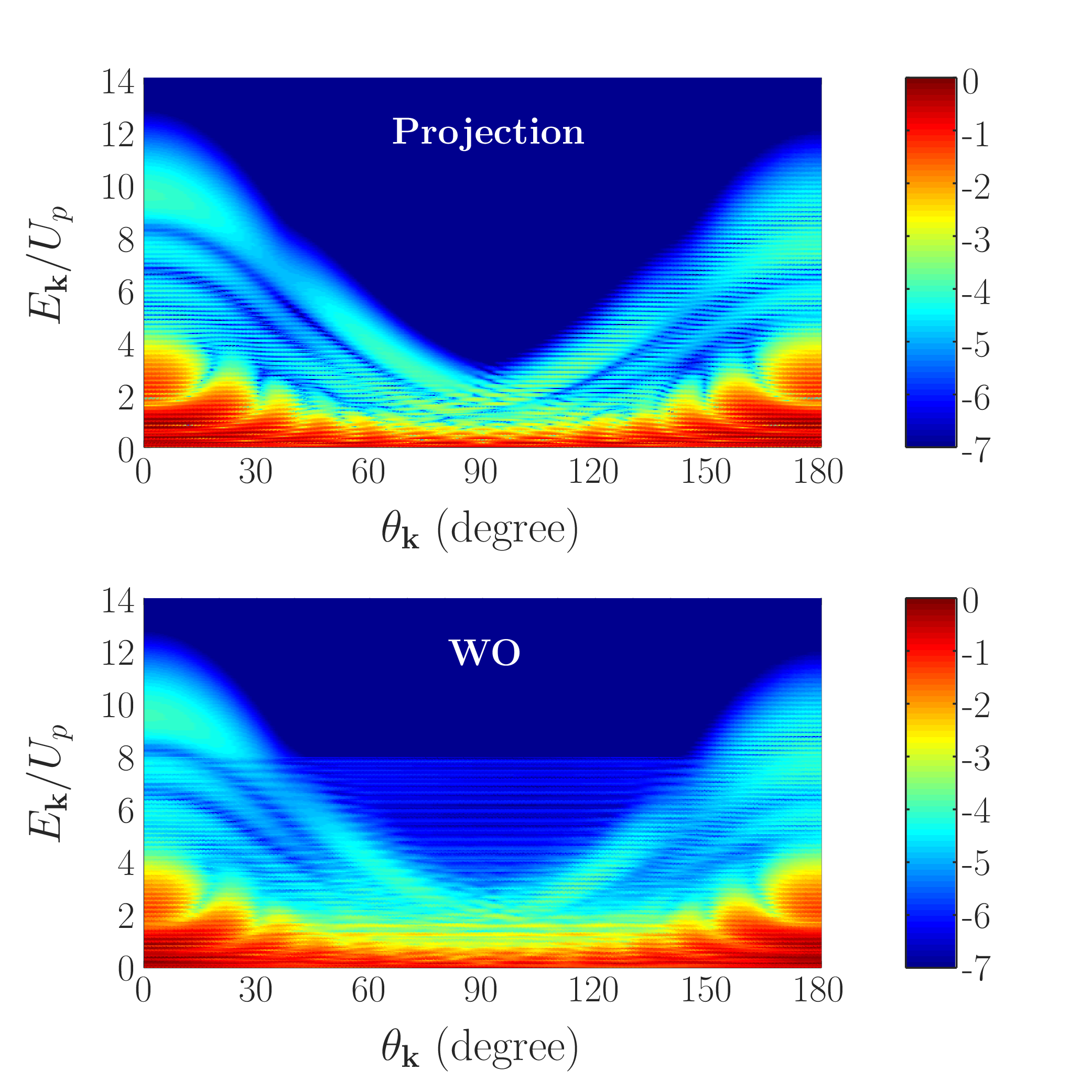}
\caption{Full PADs for the same parameters as in Fig.~\ref{results:f-_pad_vs_wop}. The upper panel shows the PAD obtained 
by projecting onto the continuum states $\Phi_\vk^{(-)}$ while the lower panel shows the PAD obtained by the WO method. The WO method gives 
additional structure for angles $\theta_\vk \in (30^{\circ}, 150^{\circ})$ and energies $E_\vk>3U_p$. }\label{results:f-_pad_full}
\end{figure}

\begin{figure}[t!]
\centering
\includegraphics[scale=0.45]{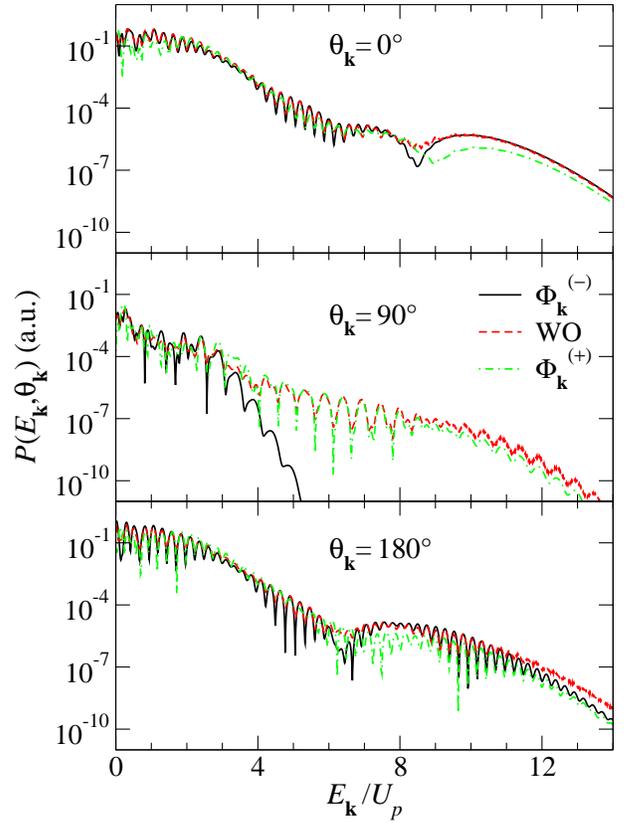}
\caption{The differential ionization probabilities of H atoms for emission of
electrons in the directions $\theta_\vk=0^{\circ}$, $90^{\circ}$, and $180^{\circ}$, as 
functions of the photoelectron energy in units of the ponderomotive energy $U_p$, for the
following laser-field parameters: $I=10^{14}~\text{W}/\text{cm}^{2}$, 
$\lambda =800~\text{nm}$, and $N_c=6$. The results are obtained by projecting 
the time-dependent wave function $\Psi(T_p)$ onto the $\Phi_{\vk}^{(-)}$ states (black solid line)
and the $\Phi_{\vk}^{(+)}$ states (green dot-dashed line) and by 
using the WO method with $\gamma=6\times 10^{-3}$ (red dashed line).}\label{results:h_pad_vs_wop}
\end{figure}

\begin{figure}[t!]
\centering
\includegraphics[scale=0.45]{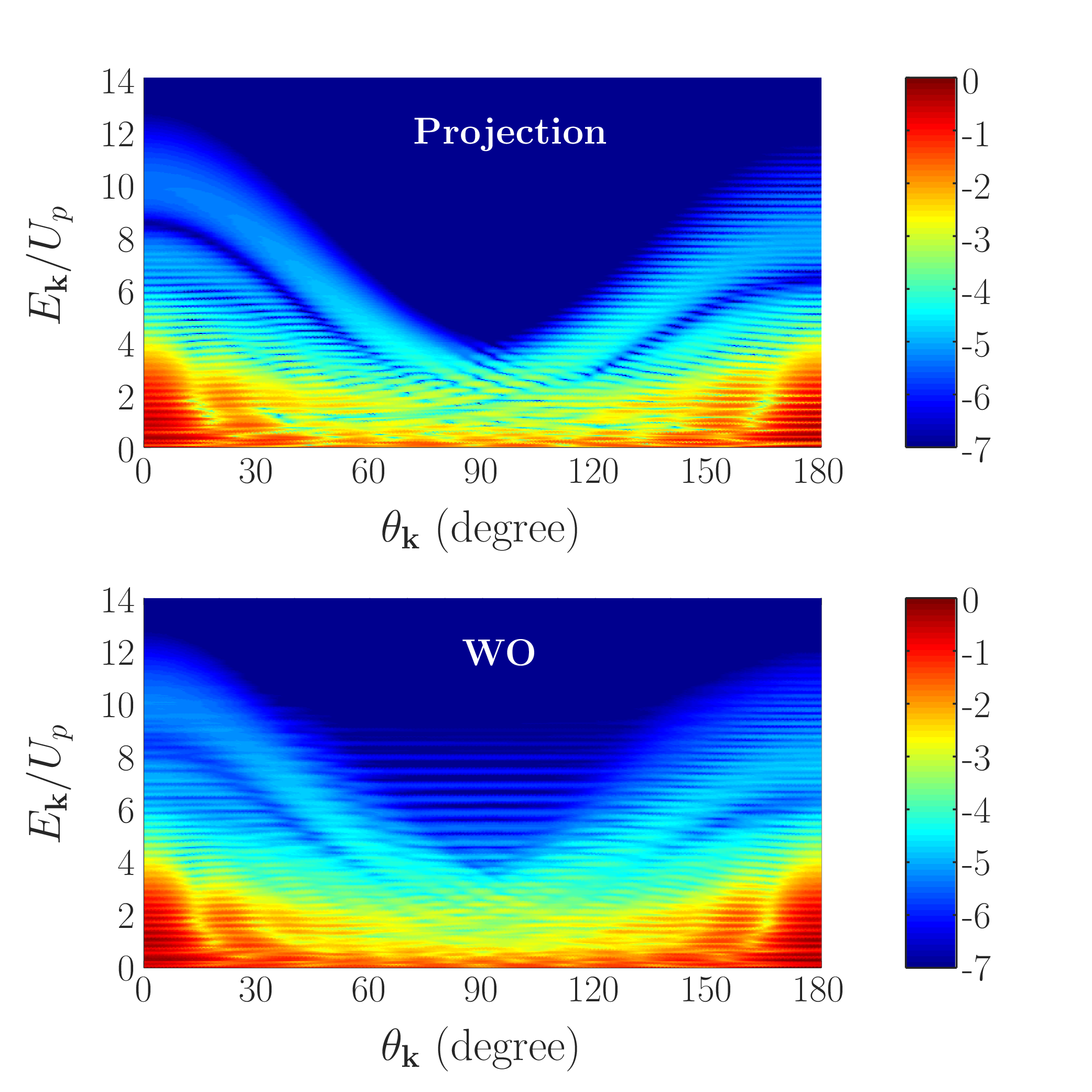}
\caption{Full PADs for the H atom and laser-field parameters as in 
Fig.~\ref{results:h_pad_vs_wop}. The upper panel shows the PAD obtained by projecting 
onto the Coulomb wave for the free particle and the lower panel shows the PAD  obtained by the WO method. The WO method gives additional 
interference structures for angles $\theta_\vk \in (30^{\circ}, 150^{\circ})$ and $E_\vk>4U_p$.}\label{results:h_pad_full}
\end{figure}

In this section we present the results for the PES obtained by the methods discussed in the previous section. We begin by comparing the spectra obtained 
using the PCS and WO methods for a short-range potential. As the target we use the fluorine
negative ion $\mathrm{F}^{-}$. Within the SAE approximation we model the corresponding potential by the Green-Sellin-Zachor potential 
with a polarization correction included \cite{GSZpot}:
\begin{equation}
V(r) = -\frac{Z}{r\left[1+H\left(e^{r/D}-1\right)\right]}-\frac{\alpha}{2\left(r^2 + r_p^2\right)^{3/2}},
\end{equation}
with $Z=9$, $D=0.6708$, $H=1.6011$, $\alpha=2.002$, and $r_{p}=1.5906$. The $2p$ ground state of F$^-$ has the electron affinity
equal to $I_p=3.404~\text{eV}$. In Fig.~\ref{results:f-_pad_vs_wop} we present the results for PAD in the directions $\theta_\vk=0^{\circ}$,
$90^{\circ}$, and $180^{\circ}$, obtained by projecting the time-dependent wave function $\Psi(T_p)$ onto continuum states
satisfying incoming boundary condition (black solid line), outgoing boundary condition (green dot-dashed line),
and using the WO method with $\gamma = 2\times 10^{-3}$ (red dashed line) for
the laser-field parameters $I=1.3\times10^{13}~\text{W}/\text{cm}^{2}$, $\lambda =1800~\text{nm}$, and $N_c=6$. 
The photoelectron energy is given in units of the ponderomotive energy $U_p=E_0^2/(4\omega^2)$. The TDSE is solved within a spherical box of 
the size $r_{\max}=2200~\text{a.u.}$ with the time step $\Delta t = 0.1~\text{a.u.}$ To achieve convergence we used
$L=40$ partial waves with $N=5000$ B-spline functions. The convergence was checked with respect to the variation of all these
parameters. The continuum states were obtained numerically in a spherical box of the size $r_0=30~\text{a.u.}$ To allow for the best 
visual comparison, the WO spectra were multiplied by a constant factor so that optimal overlap is achieved with the 
PAD given by Eq.~(\ref{pad_2}). We notice that for $\theta_\vk=0^{\circ}$ and $\theta_\vk=180^{\circ}$ these two methods produce
almost identical photoelectron spectra, in contrast to the spectrum in the perpendicular direction with respect to the polarization axis, 
i.e., for $\theta_\vk=90^{\circ}$, where we notice a significant difference. 
The WO method gives a large plateau-like annex, which extends
approximately up to $9U_p$, whereas the PAD obtained by projection onto the $\Phi_\vk^{(-)}$ states drops very quickly beyond $2U_p$.
The results obtained projecting onto the states $\Phi_\vk^{(+)}$ exhibit almost the same plateau-like annex. We will discuss this later. 
We notice here (and will again in the subsequent figures) that the calculated spectra do not observe backward-forward symmetry. This is due 
to the rather short pulse duration (recall $N_c = 6$); it can nicely be 
explained in terms of quantum orbits \cite{fewcyclerapid, rescTR}.

In Fig.~\ref{results:f-_pad_full} we present logarithmically scaled full PADs obtained either by projecting 
on the states $\Phi_\vk^{(-)}$ (upper panel) or by the WO method (lower panel). Both spectra have been 
normalized to unity and the color map covers seven orders of magnitude. 
As we can see, for small and very large angles,
these two methods produce almost identical interference structures in the PADs. However, there is a substantial difference between
the two PADs in the angular range $\theta_\vk \in (25^{\circ}, 150^{\circ})$ for $E_\vk>3U_p$.

\begin{figure}[t!]
\centering
\includegraphics[scale=0.45]{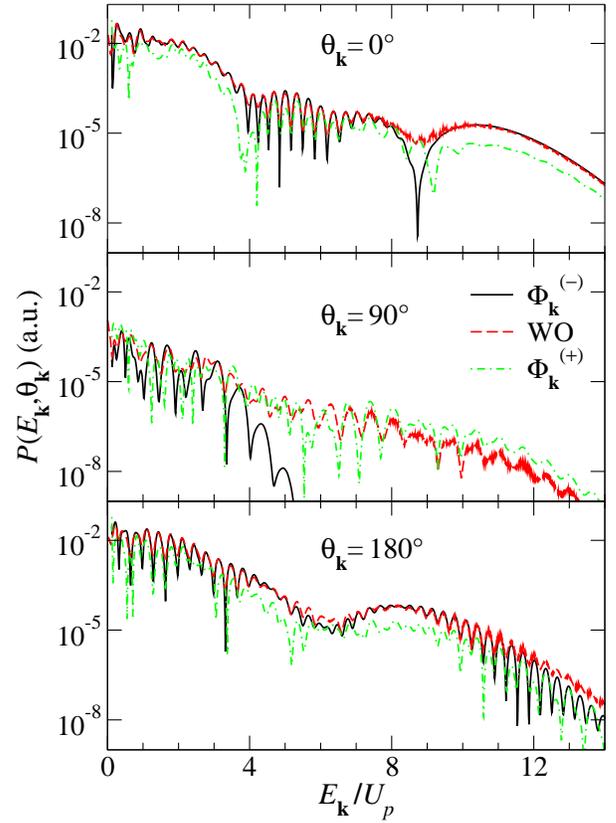}
\caption{The differential ionization probabilities of the Ar atom for emission of
electrons in the directions $\theta_\vk=0^{\circ}$, $90^{\circ}$, and $180^{\circ}$, as 
functions of the photoelectron energy in units of the ponderomotive energy $U_p$, for the
following laser-field parameters: $I=8\times 10^{13}~\text{W}/\text{cm}^{2}$, 
$\lambda =800~\text{nm}$, and $N_c=6$. The results are obtained
by projecting the time-dependent wave function $\Psi(T_{p})$ onto the 
$\Phi_{\vk}^{(-)}$ states (black solid line) and $\Phi_{\vk}^{(+)}$ states
(green dot-dashed line) and by using the WO method with $\gamma = 6\times 10^{-3}$ (red dashed line).}\label{results:ar_pad_vs_wop}
\end{figure}

Next we investigate the PAD for the hydrogen atom with its pure Coulomb potential. In Fig.~\ref{results:h_pad_vs_wop} we show the PES 
for $I=10^{14}~\text{W}/\text{cm}^{2}$, $\lambda =800~\text{nm}$, and $N_c=6$. The initial state is $1s$ ($I_{p}=13.605~\text{eV}$).
The TDSE is solved in a spherical box of the size $r_{\max}=2200~\text{a.u.}$ using $L=40$ partial 
wave and $N=5000$ B-spline functions. The time step is set to $\Delta t = 0.1~\text{a.u.}$ The spectra obtained using the WO method 
are calculated with $\gamma = 6\times 10^{-3}$. Again, we see that the WO method  as well as PCS on outgoing-boundary-condition states give a plateau-like annex in the perpendicular
direction, which is absent from the PAD obtained by projecting onto the Coulomb wave (the state $\Phi_\vk^{(-)}$). The same conclusion can be
obtained by comparing the full PADs, normalized to unity and presented in Fig.~\ref{results:h_pad_full}. In the lower panel the
PAD obtained using the WO method clearly shows additional interference structures just as in the case of $\mathrm{F}^{-}$ ions.

\begin{figure}[t!]
\centering
\includegraphics[scale=0.45]{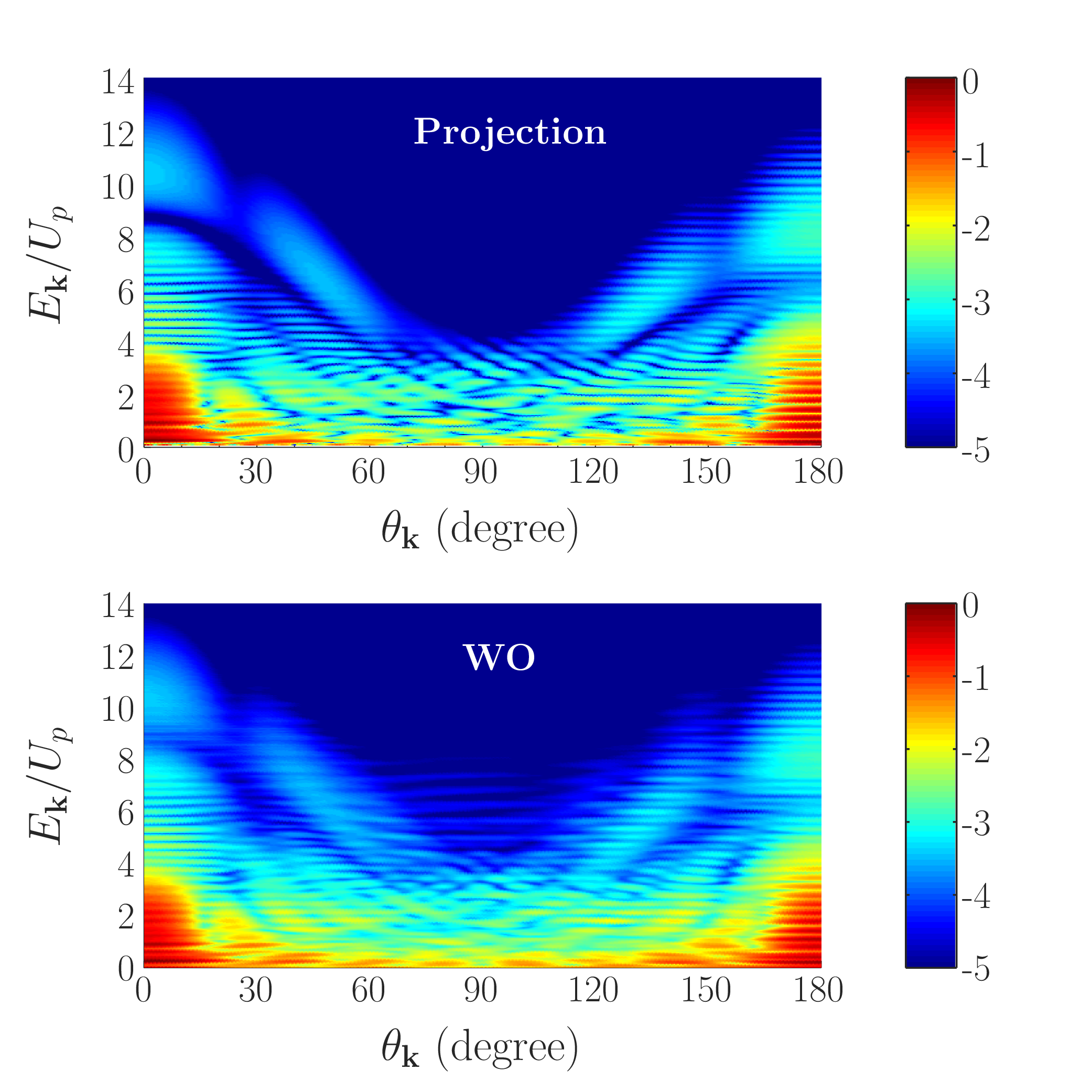}
\caption{Full PADs for the Ar atom and the same laser-field parameters as in Fig.~\ref{results:h_pad_vs_wop}. The upper panel shows the PAD 
obtained by projecting onto the continuum states $\Phi_\vk^{(-)}$ and the lower panel shows the PAD obtained by the WO method. The WO method gives 
additional structure for angles $\theta_\vk \in (30^{\circ}, 150^{\circ})$ and $E_\vk>4U_p$.} \label{results:ar_pad_full}
\end{figure}

As the last example we use modified the Coulomb potential to model the $3p$ state of the argon atom in the SAE approximation.
This potential is given by \cite{tong}
\begin{equation}
V(r) = -\frac{1+a_{1}e^{-a_{2}r}+a_{3}re^{-a_{4}r}+ a_{5}e^{-a_{6}r}}{r},\label{TongPot}
\end{equation}
with $a_{1}=16.039$, $a_{2}=2.007$, $a_{3}=-25.543$, $a_{4}=4.525$, $a_{5}=0.961$, and $a_{6}=0.443$. Using the potential (\ref{TongPot}) we calculated 
the ionization potential of the $3p$ state and obtained $I_{p}=15.774~\text{eV}$. The TDSE is solved within a spherical box of the size 
$r_{\max}=1800~\text{a.u.}$ with the time step $\Delta t = 0.05~\text{a.u.}$ Convergence is achieved with $L=40$ partial waves 
with $N=6000$ B-spline functions. The continuum states are calculated within a spherical box of the size $r_{0}=30~\text{a.u.}$ We used the
laser-field parameters $I=8\times10^{13}~\text{W}/\text{cm}^{2}$, $\lambda =800~\text{nm}$, and $N_c=6$.
The results for $\theta_\vk=0^{\circ}$, $90^{\circ}$, and $180^{\circ}$ are presented in Fig.~\ref{results:ar_pad_vs_wop}.
For $\theta_\vk=90^{\circ}$ we again notice a plateau-like structure in the spectrum obtained by the WO method and by projecting on the states  $\Phi_{\vk}^{(+)}$. This is also visible 
from the full PADs presented in Fig.~\ref{results:ar_pad_full}. 

From all these examples we can conclude that this plateau-like structure observed 
at large angles is not caused by the nature of the spherical potential $V(r)$ 
but has a different origin. Let us now explain the discrepancy between the
spectra obtained by projection on the states  $\Phi_{\vk}^{(-)}$ on the one hand 
and by projection on  $\Phi_{\vk}^{(+)}$ or by the WO method on the other, which we
 noticed in all examples presented above.
As we have already discussed, the continuum states have to satisfy the incoming
boundary condition in order to properly describe the PES. 
This boundary condition is automatically included in the continuum state (\ref{cont_st}) by the
phase factor $i^{\ell}e^{-i\Delta_{\ell}}$ for each partial wave. For a better understanding of the origin of the artificial
plateau-like annex that we see in the spectra obtained using the WO 
method, in Figs.~\ref{results:f-_pad_vs_wop}, \ref{results:h_pad_vs_wop}, and
\ref{results:ar_pad_vs_wop} we have also presented the PADs 
obtained projecting onto the continuum states $\Phi_{\vk}^{(+)}(\ver)$.
As we can see, the PAD in the direction $\theta_\vk=90^{\circ}$, calculated 
using the wrong continuum states $\Phi_{\vk}^{(+)}(\ver)$, gives the 
same artificial plateau-like structures as the WO method.
Therefore, we conclude that the effect that we see in the PADs obtained by 
the WO method is caused by the 
boundary condition satisfied by the continuum states. Since this boundary 
condition is not included or defined anywhere in the 
WO method, the energy component $\chi_{\gamma}(E_\vk,r, \Omega)$ extracted 
from the time-dependent wave function $\Psi(\ver,T_{p})$ 
is a mixture of the contributions from 
the $\Phi_{\vk}^{(-)}(\ver)$ and $\Phi_{\vk}^{(+)}(\ver)$ continuum states. That is why we 
see in the  spectrum obtained by the WO method a plateau-like structure in the 
perpendicular direction. Only the continuum states 
$\Phi_{\vk}^{(+)}(\ver)$ contribute to this spurious plateau. 
It is worth noting that another consequence of taking the wrong boundary 
condition is also visible in the  spectrum in the direction
$\theta_\vk=0^{\circ}$ for $\mathrm{Ar}$ (Fig.~\ref{results:ar_pad_vs_wop}). Namely, the 
destructive interference at approximately $8.8U_p$
is far less pronounced in the spectrum obtained by the WO method than in the spectrum 
obtained by projecting onto the states
$\Phi_{\vk}^{(-)}(\ver)$. The reason is the interplay between the two
different contributions, one that comes from the continuum 
state $\Phi_{\vk}^{(+)}(\ver)$ and the other that comes from the  $\Phi_{\vk}^{(-)}(\ver)$ 
continuum state, which is smaller by a few orders of magnitude. The same feature
we see in the spectrum for $\mathrm{F}^{-}$ for $\theta_\vk=0^{\circ}$ (Fig.~\ref{results:f-_pad_vs_wop}) 
at the kinetic energy just above $8U_p$ (it is less pronounced than in the Ar case). 

Rescattering plateaus at angles substantially off the polarization direction of the laser field like those calculated for the outgoing boundary conditions or by the WO method and exhibited in Figs.~\ref{results:f-_pad_vs_wop}--\ref{results:ar_pad_full} are difficult to understand for physical reasons. All gross features observed so far in angle-dependent above-threshold-ionization spectra have been amenable to explanation in terms of the classical three-step scenario. However, this does not allow for electron 
energies  perpendicularly to the field direction in access of about $2U_p$ \cite{rescTR,moeller14}. The reason is that within the three-step model there is no force acting on the electron in the perpendicular direction by the laser field. Hence, the perpendicular momentum has to come either from direct ionization or from rescattering. Direct ionization has a cutoff of about $2U_p$. High-energy rescattering requires that the electron return to its parent atom with high energy, and such an electron will invariably undergo additional longitudinal acceleration after the rescattering, so that its final momentum will not be emitted at right angle to the field.
 
\section{Summary and conclusions}
\label{sec:sum}
We presented a method of solving the time-dependent Schr\"{o}dinger equation (within the SAE and dipole approximations) for an atom (or
a negative ion) bound by a spherically symmetric potential and exposed to a strong laser field, by expanding the time-dependent wave function
in a basis of B-spline functions and spherical harmonics and propagating it with an appropriate algorithm. The emphasis is on the
method of extracting the angle-resolved photoelectron spectra from the time-dependent wave function. This is done by projecting the
time-dependent wave function at the end of the laser pulse onto the continuum states $\Phi_\vk$, which are solutions of the
Schr\"{o}dinger equation in the absence of the laser field (the PCS method). 
In the context of strong-laser-field ionization, the photoelectrons having
the momentum $\vk$ are observed at large distances ($r\rightarrow\infty$) in the positive time limit ($t\rightarrow +\infty$). Therefore,
it is the \textit{incoming (ingoing-wave)} solutions $\Phi_\vk^{(-)}$ that are relevant. These solutions merge with the plane-wave solutions at the time 
$t\rightarrow +\infty$: $\Phi_\vk^{(-)}(\ver,t)\rightarrow (2\pi)^{-3/2}e^{i(\vk\cdot\ver - E_\vk t)}$.

We have also presented another method of extracting the photoelectron spectra from the TDSE solutions: the window-operator method.
The WO method extracts the part of the exact solution of the TDSE at the end of the laser pulse which contributes a small interval of energies near a fixed energy $E_\vk$. The problem with this method is that it does not single out the 
contribution of the solution $\Phi_\vk^{(-)}$, but it includes an unknown linear superposition of the states $\Phi_\vk^{(-)}$ and 
$\Phi_\vk^{(+)}$. Therefore, it may lead and does lead to unphysical results, depending on the considered region of the spectrum.
By comparing the results obtained using the exact PCS method with those obtained
using the WO method for various potentials $V(r)$ we concluded that the WO method fails for an interval of the electron
emission angles around the perpendicular direction (the angle $\theta=90^\circ$ with respect to the polarization axis of the linearly 
polarized laser field). For $\theta=90^\circ$, the WO method gives a plateau-like structure, which extends up to energies
$E_\vk\sim 9U_p$, while the spectra obtained using the exact PCS method drop very fast beyond $E_\vk\sim 2-3U_p$. The full PADs show that
this unphysical structure in the spectra obtained using the WO method appears for angles $\theta_\vk \in (30^{\circ}, 150^{\circ})$ and 
energies $E_\vk>4U_p$. Furthermore, for values of the angle $\theta_\vk$ for which the results obtained using the PCS method
exhibit interference minima, the WO method smoothes out these minima, due to the spurious contribution of the states $\Phi_\vk^{(+)}$.
We have checked our results using three different type of the potentials $V(r)$: a short-range potential (F$^-$ ion), the pure Coulomb 
potential (H atom), and a modified Coulomb potential (Ar atom).

Our conclusion is that the WO method is an
approximative method that can be used to extract the photoelectron spectrum. It should be used with care since it may 
produce additional interference structures in the spectrum that have no physical significance. These additional 
structures are a consequence of the wrong boundary conditions tacitly imposed onto the continuum states by the WO method. That is why every approximative method used 
for calculating the photoelectron spectra should be tested against the exact method of projecting the time-dependent wave function onto 
continuum states satisfying incoming boundary condition.

\begin{acknowledgments}
We acknowledge support by the Alexander von Humboldt Foundation and by the Ministry for Education, Science and Youth 
Canton Sarajevo, Bosnia and Herzegovina.
\end{acknowledgments}


\begin{thebibliography}{999}


\bibitem{kulander}K. C. Kulander, Phys. Rev. A \textbf{35}, 445 (1987).

\bibitem{kulander1}K. C. Kulander, Phys. Rev. A \textbf{36}, 2726 (1987).

\bibitem{tong_gs}X.-M. Tong and S. -I. Chu, Chem. Phys. \textbf{217}, 119 (1997).

\bibitem{muller}
H. G.  Muller, Laser Phys. \textbf{9}, 138--148 (1999).

\bibitem{nurhuda}M. Nurhuda and F. H. M. Faisal, Phys. Rev. A \textbf{60}, 3125 (1999).

\bibitem{gordon}A. Gordon, C. Jirauschek, and F. X. K\"{a}rtner, 
Phys. Rev. A \textbf{73}, 042505 (2006).

\bibitem{peng}L.-Y. Peng and A. F. Starace,  J. Chem. Phys. \textbf{125}, 154311 (2006).

\bibitem{bengtsson}J. Bengtsson, E. Lindroth, 
and S. Selst\o{}, Phys. Rev. A \textbf{78}, 032502 (2008).

\bibitem{telnov}D. A. Telnov and S.-I. Chu, Phys. Rev. A \textbf{79}, 043421 (2009).


\bibitem{qprop}D. Bauer and P. Koval, Comput. Phys. Commun. \textbf{174}, 396 (2006).

\bibitem{altdse}X. Guan, C. J. Noble, O. Zatsarinny, K. Bartschat, and B. I. Schneider, Comput. Phys. Commun. \textbf{180}, 2401 (2009).

\bibitem{cltdse}C. \'{O}. Broin and L. A. A. Nikolopoulos, Comput. Phys. Commun. \textbf{185}, 1791 (2014).

\bibitem{scid-tdse}S. Patchkovskii and H. G. Muller, Comput. Phys. Commun. \textbf{199}, 153 (2016).


\bibitem{holo} 
Y. Huismans, A. Rouz\'{e}e, A. Gijsbertsen, J. H. Jungmann, 
A. S. Smolkowska, P. S. W. M. Logman, F. L\'{e}pine, C. Cauchy, S. Zamith, T. Marchenko, 
J. M. Bakker, G. Berden, B. Redlich, A. F. G. van der Meer, H. G. Muller, W. Vermin, 
K. J. Schafer, M. Spanner, M. Yu. Ivanov, O. Smirnova, D. Bauer, S. V. Popruzhenko, and M. J. J. Vrakking,
Science \textbf{331}, 61 (2011).

\bibitem{pad_exp1}X.-B. Bian, Y. Huismans, O. Smirnova, K.-J. Yuan, M. J. J. Vrakking, and A. D. Bandrauk,
Phys. Rev. A \textbf{84}, 043420 (2011).

\bibitem{pad_exp2}D. D. Hickstein, P. Ranitovic, S. Witte, X.-M. Tong, Y. Huismans, 
P. Arpin, X. Zhou, K. E. Keister, C. W. Hogle, B. Zhang, C. Ding, P. Johnsson, N. Toshima, M. J. J. Vrakking, M. M. Murnane, and H. C. Kapteyn,
Phys. Rev. Lett. \textbf{109}, 073004 (2012).


\bibitem{wop}
K. J. Schafer and K. C. Kulander, Phys. Rev. A \textbf{42}, 5794
(1990); K. J. Schafer, Comput. Phys. Commun. \textbf{63}, 427
(1991).

\bibitem{wop_app}
H. G. Muller and F. C. Kooiman, Phys. Rev. Lett. \textbf{81}, 1207 (1998);
M. J. Nandor, M. A. Walker, L. D. Van Woerkom, and H. G. Muller,
Phys. Rev. A \textbf{60}, R1771 (1999);
P. Maragakis, E. Cormier, and P. Lambropoulos,
Phys. Rev. A \textbf{60}, 4718 (1999).

\bibitem{fetic_mhati}
B. Feti\'{c} and D. B. Milo\v{s}evi\'{c}, Phys. Rev. A \textbf{99}, 043426 (2019).

\bibitem{tsurff}
A. Scrinzi, New J. Phys. \textbf{14}, 085008 (2012);
L. Tao and A. Scrinzi, New. J. Phys. \textbf{14}, 013021 (2012);
V. Mosert and D. Bauer, Comput. Phys. Commun. \textbf{207}, 452
(2016); Y. Orimo, T. Sato, and K. L. Ishikawa,
Phys. Rev. A \textbf{100}, 013419 (2019).

\bibitem{morales}
F. Morales, T. Bredtmann, 
and S. Patchkovskii, J. Phys. B \textbf{49}, 245001 (2016).

\bibitem{madsen}L. B. Madsen, L. A. A. Nikolopoulos, T. K. Kjeldsen,
and J. Fern\'{a}ndez, Phys. Rev. A \textbf{76}, 063407 (2007).


\bibitem{Bachau}H. Bachau, E. Cormier, P. Decleva, J. E. Hansen, and F. Mart\'{\i}n,
Rep. Prog. Phys. \textbf{64}, 1815 (2001).



\bibitem{roman}P. Roman, \emph{Advanced Quantum Theory: An outline of the fundamental
ideas} (Addison-Wesley, Reading, 1965), Ch. 4.2.

\bibitem{starace}A. F. Starace, in \emph{Handbuch der Physik}, edited by W. Mehlhorn 
(Springer, Berlin, 1982), Vol. 31, pp. 1--121. 

\bibitem{peng1}L.-Y. Peng and Q. Gong, Comput. Phys. Commun. \textbf{181}, 2098 (2010).

\bibitem{joachain}C. J. Joachain, \emph{Quantum Collision Theory}, 3rd ed. (North-
Holland, Amsterdam, 1983). 

\bibitem{coul90}
A. R. Barnett, in
\emph{Computational Atomic Physics. Electron and Positron Collisions
with Atoms and Ions}, edited by K. Bartschat (Springer,
Berlin, 1996), pp.  181--202.


\bibitem{deGruyter}
D. Bauer, \textit{Calculations of typical strong-field observables},
Ch.~II, pp.~43--73, in: D. Bauer (Ed.), \textit{Computational
strong-field quantum dynamics: Intense Light-Matter Interactions}
(De Gruyter Textbook, Berlin, 2016).


\bibitem{GSZpot}
P. A. Golovinsky, I. Yu Kiyan, and V. S. Rostovtsev, 
J. Phys. B \textbf{23}, 2743 (1990).

\bibitem{fewcyclerapid} 
D. B. Milo\v{s}evi\'c, G. G. Paulus, and W. Becker, Phys. Rev. A  \textbf{71}, 061404(R) (2005).

\bibitem{rescTR}
W. Becker, S. P. Goreslavski, D. B. Milo\v{s}evi\'{c}, and G. G. Paulus,
J. Phys. B \textbf{51}, 162002 (2018). 


\bibitem{tong}X.-M. Tong and C. D. Lin, J. Phys. B \textbf{38}, 2593 (2005).

\bibitem{moeller14} 
M. M\"oller, F. Meyer, A. M. Sayler, G. G. Paulus, M. F. Kling, B. E. Schmidt, W. Becker, and D. B. Milo\v{s}evi\'c,
Phys. Rev. A \textbf{90}, 023412 (2014); see also Ph. A. Korneev, S. V. Popruzhenko, S. P. Goreslavski, W. Becker, G. G. Paulus, B. Feti\'c, and D. B. Milo\v{s}evi\'c, New J. Phys. \textbf{14}, 055019 (2012).





















\end{thebibliography}
\end{document}